\documentclass[a4paper,draft]{ZbRad}
\usepackage{amssymb}
\usepackage{breakurl}
 
\usepackage[pdftex]{graphicx}
\usepackage{amscd} 

\usepackage{url}

\theoremstyle{plain}

\theoremstyle{definition}

\numberwithin{equation}{section}

\newfont{\TITf}{cmssdc10 scaled 1440}

\title[Cosmology: 100 years after A. A. Friedmann]{}

\author[Alexander F. Zakharov]{}

\begin{document}

\setcounter{page}{1}

\vspace*{40mm}

\thispagestyle{empty}

{
\TITf\setlength{\parskip}{\smallskipamount}

\begin{center}
Alexander F. Zakharov\\
Bogoliubov Laboratory for Theoretical Physics, JINR, Dubna
\bigskip\bigskip\bigskip

{COSMOLOGY: 100 YEARS AFTER A. A. FRIEDMANN}

\end{center}
}

\emph{Abstract}.

A. A. Friedmann (04.06.1888 -- 16.09.1925) proposed the first physical cosmological models in 1920s. Despite the fact that Friedmann's works were very famous soon after their publication, the study of dynamic models of the Universe in the USSR was actually banned in the 30s - 50s of the last century, and Soviet philosophers and propagandists wrote that models of the evolving Universe were invented by Lemaitre on the demand of the Roman Pope, because according to Soviet philosophers such the birth of the universe is very similar to the divine creation of the world described in the Bible. Thus, in the USSR, Friedman's works were in oblivion in since 1930s until 1960s.




\newpage\thispagestyle{empty}

\maketitle
\tableofcontents

\section{Introduction}  

In Russia, there is an opinion that the results of domestic scientists are underestimated by the international scientific community and as a result, our scientists receive significantly fewer international scientific awards than they deserve.
Earlier,  Soviet authorities  were inconsistent, as on the one hand, they demanded a minimum level of contact with foreign scientists, and on the other hand, they criticized domestic scientists for not receiving prestigious international awards. In the mid-1940s, the possibility of establishing world-class scientific and literary prizes was discussed, but these plans were not implemented \cite{Blokh_05}.

To give an example of the widespread opinion about discrimination against Russian scientists, we can recall that
the president of Soviet academy of sciences S. I. Vavilov wrote in 1947 \cite{Blokh_05}:
''According to an article in Discovery magazine, countries were ranked according to       the   number of Nobel laureates. It turned out that the USSR ended up in one of the         unenviable places, because we had only two Nobel laureates: Pavlov and Mechnikov...
Awards are given to truly worthy people. But at the same time, a large number of           equally worthy people remain ignored. And there is a system and intention in this..''
''...Popov and Marconi, Raman and Landsberg – Mandelstam..''
\footnote{Here it would be reasonable to remind that A. S. Popov died in 1905, while G. Marconi (and K. F. Braun) got their Nobel prizes for radio invention    in 1909.}
''...I can’t help but note the following: in our country, too, they are trying to hush up the   Soviet discovery.''

S. I. Vavilov expressed the same ideas in  1949:
''Among the long list of Nobel laureates since the beginning of the 20th century, there is  not a single Russian physicist. Popov did not receive the Nobel Prize, although it was     awarded to Marconi and Braun. Mendeleyev is not among the Nobel laureates,              although among them there are many people of much lesser significance in science.       P. N. Lebedev, the generally recognized best experimental physicist in the world at the  beginning of our century, did not receive the Nobel Prize. 
...Judging by the examples given, there is hardly any doubt that representatives of our        science are deliberately bypassed when awarding Nobel Prizes. This was determined to  a large extent by the fact that foreign scientists have been instilled with an arrogant         attitude towards our science and our art for centuries.''

In his Nobel prize nomination for P.L. Kapitsa in 1955 a remarkable Soviet theorist V. A. Fock wrote ''The Nobel Committee has established a custom of not giving prizes to Russian and Soviet scientists, no matter how significant their work.'' Therefore, it can be argued that the opinion of underestimating the achievements of Russian scientists was not an isolated one.

However, in order for the number of prizes to correspond to the number of significant results, the activity of the national scientific community in supporting its scientists was not significantly lower than the support of their scientists by scientific communities in other countries, as can be seen in the examples of Nobel Prizes awarded to Russian scientists (see, an interesting discussion of the issue in \cite{Blokh_05}) (see, also his paper \cite{Blokh_02}). However, as it will be shown below very often Russian scientists are not so active as researchers from other countries to promote scientific
achievements of their compatriots and nominate them for awards.
This opinion was expressed in a discussion article by the last Russian physicist, Nobel laureate V. L. Ginzburg (he was also the last Russian scientist to receive the Nobel Prize).
In particular, Ginzburg claimed \cite{Ginzburg_01} that if Russian scientists were as active as their foreign colleagues, then the ratio of American Nobel laureates to Russian ones would not be about 10--11, but about 6--7 (it would be reasonable to recall that these estimates were made by him about 25 years ago).
V. L. Ginzburg declared that it is inconsistent that  Russian scientists practically did not nominate their compatriots for the Nobel Prizes, and  they said that their outstanding achievements 
of domestic scientists were ignored by the Nobel prize Committee.

For instance, the Bureau of the Department of Physical and Mathematical Sciences of the USSR Academy of Sciences sent an extract from the minutes of the meeting dated November 1, 1955 to the Foreign Department of the USSR Academy of Sciences ''On the nomination of Soviet scientists for the Nobel Prize''.
 ''The Bureau of the department does not consider it appropriate to nominate Soviet scientists for the Nobel Prize, since, in the opinion of the bureau members, this prize cannot be considered international due to the fact that the Nobel Committee did not consider it tedious to award this prize to outstanding figures of science and culture of our country (D. And Mendeleev, L. N. Tolstoy, A. P. Chekhov, M. Gorky). The Bureau chairman  academician M. A. Lavrentiev, the secretary is Candidate of Chemical Sciences A. N. Lobachev.'' 

On the other hand, the Soviet state was atheistic and Soviet ideologists {\it a priori}   rejected  cosmological models that considered the birth and evolution of the universe.
This point of view has been consolidated in the scientific literature and the alternative view of cosmology has been rather severely criticized.
At the same time, the staff of the Central Committee of the CPSU considered it acceptable to criticize leading theoretical physicists such as Ya. B. Zeldovich, E. M. Lifshitz, L. D. Landau, 
I. E. Tamm and accuse them of incompetence as it will be shown below.
While the pressure of ideological philosophers (specialists in Marxism-Leninism) was relatively low in the field of technical applications of physics, as V. P. Vizgin aptly noted, physicists were protected by the so-called {\it nuclear shield}, the ideological pressure in cosmology was significant and affected the development of these studies.

In the USSR, there were cases of ideological pressure on scientists not only in physics and cosmology, but also in genetics, cybernetics, chemistry, and mathematics.
Some historical aspects of the development of gravity and cosmology in Russia are discussed in recent works \cite{Zakharov_25,Zakharov_25b,Zakharov_25c}.


\section{Raman vs Landsberg -- Mandelstamm}

Let us consider one case of physical discovery the combination  scattering (it is usually named the Raman scattering)
since Russian authors often they cite this example as evidence of discrimination against Russian scientists. 
A history of the discovery in Russia and India was presented in many papers, see, for instance \cite{Blokh_02b,Fabelinskii_90,Fabelinskii_03,Singh_01}. 
The combination scattering was discovered independently by G. S. Landsberg and L. I. Mandelstam on 21 February 1928 in Moscow and S. V. Raman and K. S. Krishnan on 28 February 1928 in Calcutta
(at this  time it was a part of British Raj). However, Raman and Krishnan published their paper  on 31 March 1928 in Nature \cite{Raman_28},  while the Landsberg -- Mandelstam paper was received by  on 12 July 1928 in  Zeitschrift f\"ur Physik and published in November 1928
\cite{Landsberg_28}.
Many Soviet physicists (including V. L. Ginzburg) considered this case as a discrimination of Soviet physicists since Raman received a Nobel prize in 1930, but Landsberg and Mandelstam not. 
As it is known, data on nominees and nominators becomes publicly available 50 years after the Nobel Prize is awarded. In particular, Russian scientists who nominated candidates for the Nobel Prize and were nominated for these awards become known.

\begin{table}[th]
\begin{center}
\begin{tabular}{|c c c c|} 
 \hline
 Year & Nominee & Nominator & NP Year (if there is) \\ [0.5ex] 
 \hline\hline
 1930 & Chandrasekhara Raman &  Jean Perrin &  NP in 1926\\      
      & Werner Karl Heisenberg &            &  \\  
 \hline
 1930 & Robert Wood &  Niels Bohr &  NP in 1922\\      
      &  Chandrasekhara Raman  &            &  \\  
 \hline
  1930 & Chandrasekhara Raman &  Prince Louis Victor 
  &  NP in 1929\\ 
  &       & de Broglie     &   \\  
 \hline
  1930 & Chandrasekhara Raman &  Ernest Lord Rutherford &  NP in 1908\\ 
 \hline
 1930 & Chandrasekhara Raman &  Johannes Stark &  NP in 1919\\ 
 \hline
  1930 & Chandrasekhara Raman &  Charles Wilson &  NP in 1927\\ 
 \hline
  1930 & Chandrasekhara Raman &  Eugene Bloch &  \\
    &  Robert Wood   &                        &   \\ 
     \hline
   1930 & Chandrasekhara Raman &  Maurice de Broglie &  \\ 
 \hline
 1930 & Chandrasekhara Raman & Richard Pfeiffer &  \\ 
 \hline
 1930 &Grigoriy Landsberg  &Orest Khvol’son &   \\
      & Leonid Mandelstam  & &  \\
          & (Mandelshtam) & &  \\  
      & Chandrasekhara Raman   &                                      &   \\
  \hline
 \end{tabular}
\end{center}
\caption{Raman's nominations for Nobel prize in 1930.}
\label{Raman_30}
\end{table}

\begin{table}[th]
\begin{center}
\begin{tabular}{|c c c c|} 
 \hline
 Year & Nominee & Nominator & NP Year (if there is) \\ [0.5ex] 
 \hline\hline
 1929 & Robert Wood &  Niels Bohr &  NP in 1922\\      
      &  Chandrasekhara Raman  &            &  \\  
 \hline
 1929 & Robert Wood &  Charles Fabry &  \\      
      &  Chandrasekhara Raman  &            &  \\  
 \hline
  \end{tabular}
\end{center}
\caption{Raman's nominations for Nobel prize in 1929.}
\label{Raman_29}
\end{table}

\begin{table}[th]
\begin{center}
\begin{tabular}{|c c c c|} 
 \hline
 Year & Nominee & Nominator & NP Year (if there is) \\ [0.5ex] 
 \hline\hline
  1930 &Grigoriy Landsberg  &Orest Khvol’son &   \\
      & Leonid Mandelstam  & &  \\
          & (Mandelshtam) & &  \\  
      & Chandrasekhara Raman   &                                      &   \\
  \hline
 1930 & Leonid Mandelstam & Nikolay Papaleksi &  \\
  &         (Mandelshtam)                &                 & \\
 \hline
 \end{tabular}
\end{center}
\caption{Mandelstam's nominations for Nobel prize in 1930.}
\label{Landsberg_30}
\end{table}

Table \ref{Raman_30} shows that in 1930, ten scientists nominated Raman for the Nobel Prize (six of them were Nobel laureates, four of whom received the Nobel Prize in the 1920s). Table \ref{Raman_29} shows that in 1929, Raman was nominated by two scientists, one of whom was the famous Niels Bohr. Table \ref{Landsberg_30}
shows that O. Khvol’son nominated three scientists (Landsberg, Mandelstam and Raman), while N. Papalexi nominated only  Mandelstam.
Undoubtedly, the number and status of the nominators had an impact on the selection of the Nobel Committee's experts.
Analyzing nominations for the case of the  combination  scattering discovery 
I. L. Fabelinskii ironically noted \cite{Fabelinskii_03} that ''It should be hoped that the Nobel Prize Committee takes
into consideration the value of the discovery rather than the
merits of nominators.''

\section{V. K. Frederiks: Start of GR studies in Russia}
Einstein theory of gravity (general relativity) was created in November 1915 in the result of intensive conversations between A. Einstein and D. Hilbert
\cite{Mehra_74,Earman_78,Vizgin_79,Vizgin_01,Logunov_04}.

The early development of Russian research in the field of general relativity is associated with the name of Fredericks.
Vsevolod Konstinovich Frederiks (13 April 1885 (Warsaw) – 6 June 1944 (Gorky)) was the founder of Russian studies in GR and liquid crystal physics  
\cite{Vizgin_88,Sonin_94,Sonin_95}.\footnote{The Fréedericksz (Frederiks) transition is a phase transition in liquid crystals \cite{Frederiks_27}
and this phenomenon plays a key role in liquid crystal display operation (see,  book \cite{Gennes_93} where liquid crystal physics was discussed in details).}

Vsevolod Konstinovich Frederiks was born in a noble family (his father and grandfathers were governors in Russian Empire).
In 1903 he finished gymnasium in Nizhnij Novgorod.
At the beginning of XX century Europe was a world scientific center and
Frederiks selected Switzerland (Geneve) to continue his education 
and in 1907 he finished Geneve University with specialisation in physics.
In 1909 he got PhD in physics under supervision  of famous experimentalist Charles-Eugène Guye (1866–1942).
In 1911 Frederiks  started to work an assistant at the Theoretical division of Physics Institute at Gottingen under 
Woldemar Voigt (2 September 1950, Leipzig  – 13 December 1919).\footnote{In 1887 W. Voigt found transformations which were similar to Lorentz ones \cite{Voigt_87}.
This paper was often discussed in books and papers on foundations of special relativity \cite{Miller_81,Pais_82,Heras_14}.}
 In 1914 -- 1918 Frederiks was a civil prisoner and private assistant of D. Hilbert. At this period D. Hilbert paid for Fredericks' consultations from his personal funds, 
 Hilbert requested the Gottingen University authorities to permit  Frederiks to visit to the university but this request was not granted.
 despite his request.

In 1918  Frederiks  came back in Moscow, worked in Institute of Physics and Biophysics.
In 1919 Frederiks went to Petrograd and there he was a senior physicist at State Optical Institute, a member of Atomic commission,  associate professor in Petrograd State University, professor in Pedagogical Institute.  
In 1920   Frederiks   was a lecturer at Polytechnic Institute. In  1921 
he published the first review on GR \cite{Frederiks_21}  in Soviet Physics Uspekhi (the main Soviet physical journal).
In 1999 this paper and its English translation were re-published \cite{Frederiks_21}.
In 1923 he was a senior physicist in Institute of Physics and Technology (it was founded by A. F. Ioffe in 1918). 
In 1924 Frederiks and A. A. Friedmann published of the first chapter of their joint book “Basics of GR”
(unfortunately, this book was not written due to Friedman's untimely death).
In 1926  Frederiks  was  a  Consultant to the Geological Committee.     
 In Autumn 1927 he married  Maria Dmitrievna Shostakovich (a sister of famous Russian composer D. D. Shostakovich). 
 In 1931  he was the Head of the Crystallization Laboratory in  LPTI, 
 in 1933 he was the Head of anisotropic liquid Laboratory in LPTI.
 In 1934 he   got a DSc degree (similar to Habilitation Degree) without a formal defense, nominated by the LPTI Scientific Council as a candidate for corresponding member of Soviet Academy of Science, co-editor with A. P. Afanasiev of Course on General Physics (I. K. Kikoin, Yu. B. Khariton were among authors of chapters in the book).\footnote{It is well-known that academicians I. K. Kikoin and Yu. B. Khariton  were key persons in the Soviet Atomic Project.}
 On October 20, 1936  Frederiks was arrested   as a defendant in the Pulkovo case among many others physicists and astronomers in Leningrad.
On May 23, 1937 he sentenced to 10 years in prison camps (Taishetlag). 
In 1939   he was in Orel prison,  in 1940   he was in Ukhta (Izhemlag, Komi).
 On January 6, 1944 Frederiks died. Only in 1957 relatives got an official document on his death,  where is the dash made in the place of death (but in many
 papers it was written that he died in Gorky \cite{Sonin_95} ).

\section{A. A. Friedmann and dawn of physical cosmology}
\label{sec:Friedmann}

Alexander Alexandrovich Friedmann (Friedman) 
(4(16).06.1888
 (Sankt-Pe\-ters\-burg) – 16.09.1925  (Leningrad)) was the founder of physical cosmology\footnote{An interesting book \cite{Tropp_88} was written on the centenary of this remarkable Russian scientist and the book was soon translated into English \cite{Tropp_93}.}.
We recall  some  details from his biography.
In 1897 - 1906 he was a student at the Second Sankt-Petersburg  gymnasium. 
In 1905 he wrote the first mathematical paper (published in 1907).
In 1906 - 1910  Friedmann was student at mathematical division of the faculty of physics and mathematics.
In 1910 - 1913 he left in SPb University for a preparation for a professor position.
In 1913 he passed master exams and got master degree.
In 1914 – 1916  Friedmann joined the army as a volunteer,   he served in aviation units. 
In 1918 – 1920 he was a professor of mechanics department of Perm University. 
In 1920 – 1924 he was a researcher at Atomic Commission in State Optical Institute.
In 1920 – 1924   Friedmann  was a professor at the Faculty of Physics and Mechanics of Petrograd Polytechnic Institute.
In 1920 – 1925 he was a senior physicist, head of mathematical bureau, scientific secretary and since February 1925 director of Main Geophysical Observatory.
In 1922 he  published his first cosmological paper \cite{Friedmann_22}.
In 1923 Friedmann published his book “World as space and time” (it
was the first popular presentation of GR in Russia). 
In 1923 Friedmann travelled to Germany and Norway.
In 1923  he wrote a letter to A. Einstein where he convinced Einstein that he had made a mistake in his assessment of Friedmann's cosmological solution.
In 1924  he published his second cosmological paper \cite{Friedmann_24}.
In 1924  Friedmann (with V. K. Frederiks) published the first part of their book ''Foundations of relativity theory''
In July 1925  Friedmann   and P. F. Fedoseenko made a record-breaking balloon ascent (flight at 7400 m)
In July – August 1925 Friedmann  and his wife relaxed at the Crimea cost.
On August 17, 1925  he came back to Leningrad, while his wife (N. E. Malinina) went to another town where she had duties.
Perhaps, in his last train trip Friedmann was infected by a typhus since he ate dirty fruits.   According to Friedmann himself, he probably got infected by eating an unwashed pear bought at one of the railway stations on the way from Crimea to Leningrad.  Suddenly (on September 2) he felt sick.
On  September 16, 1925  he died in hospital.
He was buried at the Smolensk Orthodox Cemetery in Leningrad.

\section{Friedmann cosmological papers}

In 1922 Friedmann published his first cosmological paper \cite{Friedmann_22} (English translation of the paper was
published in \cite{Friedmann_99}). In this paper, Friedmann showed a solution describing an expanding
universe that satisfies Einstein's equations. Einstein read the Friedmann's
paper and he published a short comment \cite{Einstein_22} (consisting of a few sentences and
one expression) where he claimed: ''The results of the work on the non-stationary
world contained in the mentioned work seem suspicious to me. In fact, it turns out
that the solution indicated in it does not satisfy the field equations.
 Friedmann was extremely disappointed by this opinion of the GR creator. 
 Friedmann consulted
with Petrograd mathematicians and sent a lengthy letter to Einstein where
Friedmann presented formulas confirming his conclusions. At the end of his letter
Friedmann requested Einstein 
 ''...Should you find the calculations presented in
my letter correct, please be so kind as to inform the editors of the Zeitschrift fur
Physik about it; perhaps in this case you will publish a correction to your statement
or provide an opportunity for a portion of this letter to be printed''.\footnote{V. Ya.  Frenkel published documents about communications between Friedmann and Einstein with active Krutkov mediation concerning the
issue \cite{Frenkel_71,Frenkel_74,Frenkel_02}.}
 Friedmann
did not receive Einstein's reply to this letter, possibly because Einstein was on a
long overseas voyage at the time. In 1923, Petrograd theorist Yuri Krutkov was
going to Germany on a scientific visit. Friedmann discussed the essence of his
arguments with Krutkov in the letter to Einstein and requested Krutkov to meet
with Einstein and convince him that Friedmann was right. In Leiden Krutkov met
Einstein at the Paul Ehrenfest house (both Krutkov and Einstein were Ehrenfest's
good friends) and Krutkov convinced Einstein that Friedmann was right. After
that Einstein wrote a short note where he mentioned that ''My criticism, as I saw
from Friedmann's letter, communicated to me by Mr. Krutkov, was based on an
error in calculations. I consider Friedman's results to be correct and shed new
light...''\cite{Einstein_23}. 
An interesting discussion of Friedmann's cosmological solutions and the historical aspects associated with them can be found in the articles and book published on the 90 and the 100th anniversary of these significant events \cite{Belenkiy_12,Belenkiy_13,Soloviev_22}.

In 1925 A. Friedman passed away untimely and after almost a century it would be reasonable to recall a sentence from the book ''Similar to Copernicus who forced the Earth to move, A. Friedman forced the Universe to expand'' 
\cite{Tropp_88,Tropp_93}.

However, Friedmann's results were not promoted and remarkable Soviet theorist  Matvei Bronstein (Sov. Physics – Uspekhi, 1931) wrote in his review of cosmological models considered in the framework of GR that “Friedmann’s cosmological solutions are half forgotten” \cite{Bronstein_31}.

\section{G. Lemaître, the Hubble law and the Big Bang concept}
Abbe Georges Henri Joseph Édouard Lemaître   (7 July 1894 – 20 June 1966) was the father of The Big Bang concept and a father of the Hubble law.
He was born in in Charleroi (Belgium).
 Lemaître signed on voluntarily on 9 August 1914 and entered in Belgium army. 
He gained his doctorate this was his licence to teach, not a PhD in 1920. 
 Lemaître got travelship to visit UK and USA in 1923.
He  visited Vesto Slipher in Arizona and Hubble at Mount Wilson in the summer of 1925 (Slipher measured redshifts for extragalactic sources and Hubble used these data
to obtain the law which was called after him).
Lemaître defended his PhD Thesis at MIT in 1927.
In 1927 he published a paper “On a homogeneous expanding universe of constant mass”, appeared in the Annals of the Scientific Society of Brussels. In the paper
\cite{Lemaitre_27}
 he derived the law (which was called later the Hubble law). Here we have to recall that in contrast to A. Friedmann who was a skillfull mathematicain Lemaître was rather educated astronomer and he knew Slipher results on redshifts  for extragalactic sources (English translation of this paper was re-published in 2013 \cite{Lemaitre_13}).

This derivation of the Hubble law was omitted by  Lemaître in the English translation published in 1931 \cite{Lemaitre_31}.
 In 2011 Mario Livio learned that Lemaître decided to omit the derivation since the law was known as the Hubble law and it was obtained from observations
 \cite{Livio_11b}.\footnote{In 2018 International Astronomical Union decided to name the Hubble law as the Hubble – Lemaitre law.}

In his paper \cite{Lemaitre_27} Lemaître did not quoted Friedmann's papers on the subject perhaps due to the fact that  Lemaître ddi not read papers written in German.
In spite of that according to J. P. Luminet's opinion the Big Bang concept had two fathers: Russian Alexander Friedmann and Belgian Georges Lemaitre \cite{Luminet_23}.
''...André Deprit (a former student of Lemaître) gives a more picturesque and slightly different version of this encounter. In particular, he states that Lemaître did not know German, which may explain why the Belgian scholar did not cite Friedmann's earlier work in his 1927 article   \cite{Luminet_23}.''

 In 1927 Lemaître had conversations with Einstein \cite{Mitton_17}:
''In October 1927, Einstein participated at the 5th Solvay Congress in Brussels, where the main topic was quantum theory.  Lemaître was not invited to those discussions, but Louvain is only 20 km from the capital, so during a break in the closed sessions he caught Einstein’s attention, and the two took a stroll in the Parc Leopold. Einstein commented favourably on Lemaitre’s mathematical competence, although he rejected the notion of an expanding universe as an abomination. And it was Einstein who, at this encounter, directed 
 Lemaître’s attention to Friedman’s work.    Lemaître found this news unsettling, but his confidence in Slipher’s data compelled him to continue to work on expansion as the key to interpreting the data on ''nebular velocities''.'' 


''During his meeting with Einstein, Lemaître had been informed about the paper of Friedmann of 1922, of which he was not previously aware. This was probably related to the fact that Lemaître was more interested in the astronomical literature than to articles of physical theory, especially when those were written in German, a language that he had not mastered as well as English.In the future, referring in his publications to the equation describing the variation of the radius of a spherical, homogeneous and isotropic universe, mentioned in his famous 1927 paper, he would always call it the ''Friedmann equation'', without trying to emphasize his own contribution'' \cite{Lambert_15}. 

About the introduction of the Primeval Atom concept (a Hot Universe model) an outstanding cosmologist James Peebles (he got Nobel prize in physics in 2019) wrote \cite{Peebles_84}: ''Physical scientists have a healthy attitude toward the history of their subject: by and large we ignore it. But it is good to pause now and then and consider the careers of those who through a combination of the right talent at the propitious time have had an exceptional influence on the progress of science. As I have noted on several occasions it seems to me that Georges Lemaître played a unique and remarkable role in setting out the program of research we now call physical cosmology.''


In 1931   Lemaître published a paper where he introduced a hot Universe model \cite{Lemaitre_31b}
(the paper was re-published almost after 80 years \cite{Lemaitre_11}.
In 1934    Lemaître   evaluated a background temperature in a few K  using his primeval atom model \cite{Lemaitre_34} (as it was noted by J.-P. Luminet it  was an important feature of a Hot Universe model \cite{Luminet_11}).
In 1946,  Lemaître’s book L’Hypothesis de l’Atome Primitif, Essai de Cosmologie had described his ideas at greater length, paying more attention to concepts of the quantum world
\cite{Lemaitre_46} (see, also English translation \cite{Lemaitre_46e}). A contemporary view on Lemaître’s famous Primeval Atom Hypothesis (PAH) was presented
in \cite{Lambert_25}.
In 1948, Ralph Alpher and Robert Herman (both were Gamow's students) published  an estimate of 5 K for the temperature of the cosmic microwave background for hot Universe model \cite{Alpher_48}.


In 1933 at Caltech after the   Lemaître’s lecture A. Einstein said: “This is the most beautiful and satisfactory explanation of creation to which I have ever listened”. 
Lemaître was invited to
deliver a lecture at the Mount Wilson Observatory, where E. Hubble worked. A.~Einstein followed this Lemaître's lecture. When journalists asked Einstein about
his impression of Lemaître's cosmological model, Einstein replied, ''This is the most
beautiful and satisfactory explanation of creation to which I have ever listened!''\footnote{D. Lambert, Einstein and Lemaître : Two friends, two cosmologies..., 
\url{https://inters.org/einstein-lemaitre}.
}
This Einstein’s opinion was widely distributed through mass media. These circumstances had a negative impact on the development of cosmological studies in USSR for around thirty years.

After that journalist Dunkan Aikman published interview with Lemaître with
the title {\it Lemaître follows two paths to truth: The famous scientist, who is also a
priest, tells why he ends no conflict between science and religion} (the interview was
published in New York Times on February 19, 1933)\footnote{https://www.nytimes.com/1933/02/19/archives/lemaitre-follows-two-paths-to-truth-the-famous-physicist-who-is.html}. This article was extremely
popular over the world. Because the theses that religion can open the way to
truth, and that religion does not contradict science, were unacceptable to Soviet
ideologists. They found a way out of this situation, just as it was later done
for genetics and cybernetics. Soviet ideologists came to the conclusion that the
approach to constructing cosmological models based on general relativity should
be declared pseudoscientific, and the Soviet scientists who worked on this topic
should be declared pseudoscientists and admirers of Western bourgeois science.
The Pulkovo astronomer M.~Eigenson wrote a book 
\cite{Eigenson_36}\footnote{According to Shklovsky's memoirs \cite{Shklovsky_91}, Eigenson was the secretary of the Communist party organization of the Pulkovo Observatory and was involved in the development of the Pulkovo case in 1936.}
where such a point of view on cosmological models was formulated. Thus, for about 30 years there have been two points of view on cosmology. According to the Soviet point of view, the Universe as a whole is infinite in time and space (the entire Universe does not expand)
 and a Friedmann expansion is possible only in a local area, while most Western astronomers believed that Friedman's solution was applicable to the entire Universe. We do not discuss so called steady state Universe model proposed by H. Bondi, T. Gold, F. Hoyle, G. Burbidge and J. Narlikar.
Thus, a division occurred in cosmology: Western cosmology, which considered
realistic dynamic models of the Universe, and Soviet cosmology, which asserted
that the Universe is infinite in time and space. It was done in contradiction to
a famous Chekhov's statement that there is no national science as there is no
national multiplication table. So, Soviet ideologists declared that the Universe
must be infinite in space and in time in contrast to the Friedmann -- Lemaître point of
view on the subject.

During this period, it was believed that Soviet cosmology
claimed that the Universe was infinite in time and space, and
only such a concept corresponded to the then prevailing philosophy
of dialectical materialism in the Soviet Union; otherwise,
scientists were recognized as representatives of idealism  \cite{Graham_87,Wetter_58}.
In particular, regarding the cosmological solutions of
A.A. Friedmann, the Great Soviet Encyclopedia stated that of
course, these  [A.A. Friedmann]   solutions could not be rightfully
considered as the basis for models of the entire Universe \cite{Zelmanov_55}. This article identifies Eigenson's book \cite{Eigenson_36} as the main source of information on cosmology.

Analysing Soviet declarations G. Wetter noted \cite{Wetter_58}: 
  ''From the above account of the basic themes of Soviet cosmology and cosmogony it will be seen that they have two main objects particularly in view: the first is to banish from modern astronomy all theories which in any way involve the assumption that the universe had a beginning in time or is spatially finite, and thereby encourage the idea of a creation of the world. In addition to this we find, in the field of cosmogony, the same fear of geocentricity.''

\section{Soviet atheistic ideology and physical cosmology}

%
Soviet ideologists carefully observed that Soviet scientists, in particular physicists, did not deviate from the views of Marxist-Leninist philosophy that space and time should be infinite in time and space.

As an example, we can give an example of a memo from the Department of Science, Universities, and Schools of the Central Committee of the Communist Party of the Soviet Union.
This letter was written by administrators who criticized leading Soviet physicists and accused them of being incompetent theoretical physicists. Three of these physicists later became Nobel laureates (I. E. Tamm, L. D. Landau, and V. L.  Ginzburg), and Academician Ya. B. Zeldovich was widely recognized internationally and was awarded three Hero of Socialist Labor awards for his work on the atomic project.\footnote{By analyzing the citation, Scopus has identified who are part of the 2\% of active global scientists. 
The Scopus published a list of the active scientists, Ginzburg and Zeldovich are still among them, despite the fact that Zeldovich passed away in 1987 and Ginzburg in 2009. 
All world active scientists (according to Scopus criteria) are given at \url{https://elsevier.digitalcommonsdata.com/datasets/btchxktzyw/8}, while 
Russian scientists (around 1000 names) are selected in file \url{https://disk.yandex.ru/i/i51WY6MgZ9QbPA}.}

''{The Head of the Department of Science, Universities, and Schools of the Central Committee of the Communist Party of the Soviet Union, V. A. Kirillin, Deputy Head, N. I. Glagolev, and Instructor A. S. Monin at the Central Committee of the Communist Party of the Soviet Union wrote the memo, dated January 16, 1956}.
{ On November 30 - December 1, 1955 the Department of Physical and Mathematical Sciences of the USSR Academy of Sciences held a scientific session dedicated to the 50th anniversary of the theory of relativity. There were serious shortcomings in the preparation and conduct of the session. The program of the session, prepared by the organizing committee, which included academicians Tamm and Landau, corresponding member of the USSR Academy of Sciences Ginzburg and Professor Lifshitz, was unsatisfactory, as it included reports by Academician Landau, corresponding member of the USSR Academy of Sciences Ginzburg and Professor Lifshitz, who do not work in the field of relativity theory} [it was true -- AZ] { and are known for their nihilistic attitude to the development of methodological issues of this theory}...
{ After requesting abstracts from some of the session participants, the organizing committee did not request abstracts from Comrade Lifshitz. No transcripts were kept at the session meeting. These omissions turned out to be serious, as there were serious ideological errors in the comrade Lifshitz's report. This report, devoted to a review of research on relativistic cosmology, was a significant propaganda of the idealistic "theory of the expanding universe." This "theory", which is an illegal extension of non-stationary solutions of Einstein's gravitational equations to the universe as a whole, was {\bf built by Abbé Lemaitre on the direct order of the Pope}} [highlighted by the author -- AZ].
{ ...Ignoring the work on the hierarchical structure of the distribution of matter in the universe, in particular, on the structure of metagalaxy, T. Lifshitz argued that matter is distributed in the universe with a constant average density, as allowed in the "theory of the expanding universe", and stated that the value of this density can determine whether the "expanding universe" has a finite or an infinite volume. The assessment given by  comrade Lifshits turned out to be just on the edge between these possibilities. Next, comrade Lifshitz calculated the age of the universe (calling it "characteristic time"). The coincidence of the obtained "age" (several billion years) with a certain radioactive method of the age of the Earth's geological rocks, indicating the meaninglessness of such an estimate not only for the metagalaxy, but also for individual galaxies,  comrade Lifshitz declared non-accidental and claimed that this coincidence allegedly confirms the "theory of the expanding universe." This statement was supported by Zeldovich, a corresponding member of the USSR Academy of Sciences, who said that not only the geological rocks of the Earth, but also chemical elements in general, have the same age. At the same time, the expression "the time from the moment when the trigger was pulled" was used. When presenting these issues,  comrade Lifshits made no reservations about the limited applicability of the "models" under consideration and their ideological meaning...}.
{ We would consider it necessary to point out to the Presidium of the USSR Academy of Sciences that insufficient attention is being paid to the preparation and holding of meetings at the Department of Physical and Mathematical Sciences, and also suggest that the Presidium of the USSR Academy of Sciences organize, within a month, a discussion of the ideological errors made in T. Lifshitz's report at the bureau of the Department of Physical and Mathematical Sciences. We request for your consent...}
{ On the left side of the first sheet is the entry: "Agree. Instruct Comrade Kirillin to do this in some form... (further illegible).
D. Shepilov. 26.1.56." Below are the signatures of P. Pospelov, M. Suslov, A. Aristov, N. Mukhitdinov.}'' \cite{Blokh_05}.

\label{sec:Gamow}
\section{G. A. Gamow and his hot Universe model}
In 1940s G. Gamow proposed the hot Universe model, calculated primodial nucleosynthesis and predicted an existence of  Cosmic Microwave Background (CMB)  radiation. 
As it was noted  his students evaluated a temperature of   CMB
radiation which should have now temperature around 5 K. In consequent calculations
the CMB temperature was slightly different, but always it was a few
kelvins. It would be reasonable to note that similar estimates for background radion
temperature were done by Lemaître as we noted earlier.
How to estimate temperature without analyzing primary nucleosynthesis is shown in an interesting article by A. D. Chernin \cite{Chernin_94}, written for Gamow's 90th birthday.

As a young man, Gamow was considered one of the most successful Soviet physicists and was elected a corresponding member of the USSR Academy of Sciences at the age of 28. However, after he left the USSR permanently in 1932, he was perceived by the Soviet leadership as an unloyal former citizen of the country, and as a result, Soviet scientists preferred not to mention Gamow's work.
G. Gamow said on communications with Soviet Physicists on one of his last interview:
''So a couple of years ago ... You see, my situation with Russian scientists is that physicists and astronomers know that I am persona non grata, and they are afraid to write to me, and I don't want to write to them because I bring them into trouble. But biologists don't. A Russian name, well, there are many Russian names. So a couple of years ago I got from Luchnik some reprints. He was interested in this coding problem, and I sent him my reprints and letter. And this is my stationery, which probably you have seen''.\footnote{https://www.aip.org/history-programs/niels-bohr-library/oral-histories/4325.}

\label{sec:Shmaonov}
\section{T. A. Shmaonov and the CMB discovery in Pulkovo}

The CMB (cosmic microwave background) radiation or relic radiation in Russian literature is one of
the key signatures of the hot Universe model developed by G. Gamow in the 1940s
and 1950s. The CMB radiation was discovered by T. Shmaonov at the Pulkovo
Observatory several years before A. Penzias and R. Wilson \cite{Shmaonov_57}.
At that time, Shmaonov was a post-graduate student of Khaikin at the Pulkovo Astronomical Observatory. When Shmaonov asked Khaikin about a possible interpretation of the result, Khaykin replied that he did not have an interpretation, but that the result should be published, which was done, and the corresponding article was sent to a non-astronomical journal (which was not translated into English at the time). Khaikin was one of the most brilliant and well-educated representatives of the L.I. Mandelstam school
(a scientific biography of this remarkable scientist was recently published \cite{Yakuta_21}). Gamow and his works were well-known to the Mandelstam school, but  citations of Gamow's papers were  not wellcome, so Khaikin chose not to inform the scientific community that his student had confirmed Gamow's predictions. 
As we have already said, even the use of models of the expanding universe (where the birth and evolution of the universe are discussed) was considered religious propaganda and such discussions were severely criticized.
Khaikin selected the option when neither he nor his student would have any problems.

The discovery of Shmaonov  unfortunately was quoted very rarely, in spite of the fact, that
it is known in the world scientific literature, see, for example, Trimble's article \cite{Trimble_06} or the text in the book
where it was written  \cite{Peebles_09}:
''Shmaonov described how, in the middle of the 1950s, he had been doing postgraduate research in the group of the well-known Soviet radio astronomers S. Khaikin and N. Kaidanovsky: he was measuring radio waves coming from space at a wavelength of 3.2 cm. Measurements were done with a horn antenna similar to that used many years later by Penzias and Wilson.  Shmaonov carefully studied possible sources of noise. Of course, his instrument could not have been as sensitive as those with which the American astronomers worked in the 1960s. Results obtained by Shmaonov were reported in 1957 in his PhD Thesis and published in a paper (Shmaonov 1957) in the Soviet journal Pribory i Tekhnika Eksperimenta (Instruments and Experimental Methods). The conclusion of the measurements was: “The absolute effective temperature of radiation background ... appears to be $4 \pm 3$~K.'' 
Shmaonov emphasized the independence of the intensity of radiation on direction and time.'' 

\section{Official recognition of Friedmann achievements}

In 1963 Soviet physicists and cosmologists were granted the right to mention the Friedmann cosmological model in a positive light and to consider the physical and astronomical implications of using this model. This event was preceded by the active work of physicists. In June 1963 a scientific community celebrated
75 years since A. A. Friedmann's birthday and Department of Physics and Mathematics of Soviet Academy of Sciences organized a specail session dedicated to this event.
 Before the session, at the general meeting of Soviet Academy of Sciences on February 6, 1962  P. L. Kapitsa  said \cite{Kapitsa_62}:
''What has been said about physics can be applied to other areas of the natural sciences. The separation of theory from experiment, experience, and practice damages, first of all, the theory itself. I would like to say that the separation from experience and from life also occurred among philosophers who study the philosophical problems of natural science.
Here is another example that shows what insufficient understanding and knowledge of physical experiments leads to. Many still remember freshly how a number of philosophers, dogmatically applying the method of dialectics, proved the inconsistency of the theory of relativity. The greatest criticism from philosophers was subjected to the conclusion of the theory of relativity that energy is equivalent to mass multiplied by the square of the speed of light ($E = mc^2$). Physicists have long verified this law of Einstein in experiments with elementary particles. To understand these experiments, deep knowledge of modern physics was required, which some philosophers did not have. And so physicists carried out nuclear reactions and tested Einstein's law not on individual atoms, but on the scale of an atomic bomb. Physicists would be good if they followed the conclusions of some philosophers and stopped working on the problem of applying the theory of relativity to nuclear physics! In what position would physicists have put the country if they had not been prepared for the practical use of the achievements of nuclear physics?
This shows that the application of dialectics in the field of natural sciences requires an exceptionally deep knowledge of experimental facts and their theoretical generalization. Without this, dialectics itself cannot provide a solution to the problem. It is, so to speak, a Stradivarius violin, the most perfect of violins, but in order to play it, one must be a musician and know music. Without this, it will be as out of tune as an ordinary violin.''

P. L. Kapitsa said at the session \cite{Kapitsa_63}:  ''Friedmann made one of the most significant theoretical discoveries in astronomy—he predicted the expansion of the Universe. From Friedmann's solution of Einstein's cosmological equations, it followed that the radius of curvature of our world could change over time. A few years after Friedmann's work was published, the American astronomer Hubble discovered the recession of galaxies—a consequence of the expansion of the Universe. Thus, Friedmann "at the tip of his pen" discovered an amazing phenomenon of cosmic scale... 
Friedmann did not live to see his calculations confirmed by direct observation. But we now know that he was right. And we are obliged to give a fair assessment of the remarkable result of this scientist. Friedmann's name has been undeservedly forgotten until now. This is unfair and this needs to be corrected. We must perpetuate this name. After all, Friedmann is one of the pioneers of Soviet physics, a scientist who made a great contribution to domestic and world science. It is necessary to publish a collection of all his works and publish his biography.''

In 1963 Ya. B.  Zeldovich  
delivered a talk and after that he published paper \cite{Zeldovich_64}, where he spoke about the significance of Friedmann's work for cosmology, and in addition, in 1963 Zeldovich began to mention Gamow's work on the hot Universe model for the first time in Soviet scientific literature. Undoubtedly, the courage and boldness of Zeldovich for such actions were necessary for this, since Gamow was persona {\it non grata} for the official Soviet science of that time. 

Academician V. Fock reminded in his talk at the session \cite{Fock_64}:''Ale\-xander Alexandrovich Friedman and Vsevolod Konstantinovich Fredericks, being professors at Petrograd (now Leningrad) University, were the first to introduce Russian physicists working in Petrograd to Einstein's recently created theory of gravity. It was at the very beginning of the twenties, when the blockade of Soviet Russia had just been broken and scientific literature began to arrive from abroad. A seminar was held at the university's Physics Institute, where, among others, reports on Einstein's theory were presented. The seminar participants were professors and senior students (there were few of them then). The main speakers on the theory of relativity were V. K. Fredericks and A. A. Friedmann, but sometimes Y. A. Krutkov, V. R. Bursian and others spoke. I vividly remember the reports of Fredericks and Friedmann. The style of these reports was different: Fredericks deeply understood the physical side of theory, but did not like mathematical calculations, while Friedman focused not on physics, but on mathematics. He strove for mathematical rigor and attached great importance to a complete and accurate formulation of the initial assumptions. The discussions between Fredericks and Friedmann were very interesting.''  

In the Soviet Union, one of the main scientific disciplines was Marxist-Leninist philosophy. Despite the fact that models of an expanding universe were  allowed
for their considerations by physicists and cosmologists 
 since 1963, Soviet philosophers considered cosmology to be their own field of expertise. 
There was no other philosophy in the country, or rather any other point of view on philosophy was, by definition, erroneous. In all Soviet institutes and universities, this subject was taught to all students of all specialties for one year.
Textbooks and books on Marxist-Leninist philosophy often postulated positions that did not correspond to scientific ideas.
The contents of these textbooks were only to be memorized, not critically evaluated.
The Marxist-Leninist philosophy was one of the main Humanities until the end of Soviet Union.
The number of such textbooks published every year was several hundred thousand. 
The last Russian Nobel laureate in physics, V. L. Ginzburg, who received the award in 2003, criticized this practice of postulating claims that were not supported by observations or experiments.
In particular, Ginzburg criticized the following cosmological statements
done by Soviet philosophers ''space and time are limitless and infinite'';
''the space of the Universe is not only limitless, but also infinite'' (page 68 in  \cite{Konstantinov_73});
''the infinity of space is the infinity of the volume of the entire uncountable set of material bodies in the Universe''(page 78 in  \cite{Konstantinov_73}).
Discussing the statements Ginzburg noted \cite{Ginzburg_81}, that in this case the closed cosmological model is rejected without any reason or observational evidence.

In the next edition of the textbook \cite{Konstantinov_79} (which was written by 16 authors, one of whom was an academician, 4 were corresponding members of the USSR Academy of Sciences, the
rest were doctors of Philosophy, and this book was published in an edition of 300,000 copies)
the above quotes about the infinity of space disappeared and the following statements appeared
''One of the universal properties of space and time is their infinity...''
''All assumptions of the finiteness of time inevitably lead to religious
views about the creation of the world and time by God, an assumption
that has been completely disproved by all the facts of science and
practice...''
''Matter is infinite in its spatial forms of existence...''
Ginzburg also criticized these statements.
He wrote \cite{Ginzburg_80} ''Questions about the finiteness or infinity of the volume of this Universe, laws of
its evolution in time, and  similar considerations are not philosophical questions
and must be decided in the light of specific astronomical observations and modem
physics.''


\section{Conclusion}

From 1930s until the beginning of 1960s dynamical cosmological models were banned.

In September 2025 a scientific community marked the 100th anniversary of the death of A. A. Friedmann and the 110th anniversary of the GR creation.  The above are some results in the development of gravity and cosmology that are not covered in great detail in the Russian literature, for example, when discussing the Big Bang model in Russian literature, Lemaître's contribution is often forgotten, and the CMB discovery of Shmaonov  are very rarely mentioned.
Unfortunately, very often our compatriots are not active to promote achievements of other researchers (including our domestic achievements). 
In these circumstances people remember the M. I. Monastyrsky's dictum \cite{Monastyrsky_09} ''...By the way, we (in Russia) are very fond of lamenting the underestimation of Russian scientists in the West. Russian  (Soviet) scientists are the ones who hinder the recognition of Russian (Soviet) scientists the most, however, a careful analysis of the real facts shows...''

Many years ago the President of Soviet Academy of Sciences S. I. Vavilov expressed a similar opinion ''I cannot help but note the following: unfortunately, in our country, the Soviet discovery is often ignored'' \cite{Blokh_02b}.
This sad tradition was not only in Soviet Union but also in Russian Empire.
Mendeleyev is one of the most famous Russian scientists, but he was not a full member of the St. Petersburg Academy of Sciences.
When in 1880 the incident with the non-election of D.I. Mendeleyev as an academician occurred, Russian Prime Minister S.Yu. Witte wrote to the president of the academy K.K. Romanov about 
D.~I.~Men\-deleyev as a person representing the type of amazing Russian scientist: ''If he were a Frenchman, a German, an Englishman - he would have long ago been a member of the highest scientific national institution. His name is known to the whole world... But I know that there will always come a moment when the highest feelings, the highest intentions push aside the lower ones and justice is given to each according to his merits. But will this moment come for old Dmitri Ivanovich Mendeleyev?''
\cite{Blokh_05}.

Unfortunately, the discovery of cosmic microwave background radiation by Shma\-onov is very rarely mentioned in both domestic and foreign literature. For a long time, the remarkable achievements of Friedmann and Gamow related to cosmology research did not receive due attention and support in our country. In Russia,  Lemaître's works on the development of the Big Bang theory (the hot Universe model) were underestimated. In Soviet (atheistic) times, some prejudice among Russian scientists was due to the fact that Lemaître was a Catholic priest.

Undoubtedly, there were scientists in the USSR and in Russia whose level
of achievement corresponded to the level of Nobel laureates, but they did not live long, and, as Ginzburg repeatedly said, to receive the Nobel Prize, you need to be long-lived.
In addition, there are many names of famous foreign scientists who did not receive the Nobel Prize. It can be argued that Russian scientists are much less active in promoting the achievements of their compatriots, and it is not possible to give a convincing example of discrimination against Russian scientists.

Serious administrative pressure and underestimation of value of scientists and inventors led to the so called  ''brain drain'' and  serious economical and financial losses. 
  Several years ago the dean of economical faculty at Moscow State University A. Auzan said: ''Russian losses from                  V. K. Zvorykin immigration are equivalent to 10 Russian GDP. From the Sergey Brin immigration are around 5 GDP''.

\section*{Acknowledgements}

The author thanks the organizers of the 
International Conference on the Occasion of Branko Dragovich 80th Birthday for their invitation to present my contribution as an invited talk at this event and their attention to the research presented in the paper and  V.O. Soloviev, E.E. Donets,  P. Jovanovi\'c, V. Borka Jovanovi\'c,  D. Borka
  for useful discussions.

\end{document}